\documentclass[12pt]{amsart}

\newcommand{\R}{\mathbb R}
\newcommand{\HH}{\mathcal H}

\newcommand{\eqdef}{\stackrel{\text{def}}{=}}

\begin{document}

\title{Quantization on space-like surfaces}
\author{A. V. Stoyanovsky}
\email{stoyan@mccme.ru}
\address{Russian State University of Humanities}
\begin{abstract}
We give a mathematical definition of dynamical evolution in quantum field theory, including evolution on
space-like surfaces, and show its relationship with the axiomatic and perturbative
approaches to QFT.
\end{abstract}
\maketitle

\section{Introduction}

In this note we give a mathematical definition of dynamical evolution in quantum field theory. Some kind of such
definition has been given in Wightman's axioms of QFT. Wightman considered a Hilbert space of states with action of
Poincare group. A more general although less rigorous approach has been first proposed by Tomonaga [1] and
Schwinger [2]. They proposed to consider evolution of quantum states in Hilbert space
corresponding to various space-like surfaces in
space-time. Later this approach has been discussed by Dirac [3] who called it ``quantization on space-like surfaces''.
Although Wightman axioms, extended further by Lehmann, Symanzik, and Zimmermann (WLSZ) (see, for example, [4]),
give an approach consistent for
many purposes, it is clear that a more deep consideration should involve quantization on space-like surfaces. This is
necessary, in particular, for relation with general relativity, and for generalization of QFT
to pseudo-Riemannian and more general manifolds.

In 1998 Torre and Varadarajan showed [5] that the approach to dynamical evolution of quantum states
in Hilbert space corresponding to space-like surfaces meets serious difficulties already in the case of
a free scalar field. Namely, they have shown that linear canonical transformations of the phase space of a free
Klein--Gordon field corresponding to evolution of solutions of the Klein--Gordon equation between two curved Cauchy
space-like surfaces, cannot be in general implemented by unitary operators in the Fock space. In the present paper we show
how to overcome this difficulty, and define quantization on space-like surfaces mathematically. The key point in
our approach is the use of divergent unitary transformation called the Faddeyev transform [6,7]. Using this transform
we define quantization on space-like surfaces first in perturbative QFT, and then show how to provide a mathematical
definition extending the WLSZ axioms.

The author is obliged to V. P. Maslov for illuminating discussions.

\section{The Faddeyev transform and quantization on space-like surfaces in QFT}

What is the relation between perturbative and axiomatic QFT? In particular, how to recover the Hilbert space $\HH$ with
the action of Poincare group from renormalized perturbation series? The answer is given using the so called Faddeyev
transform [6,7]. Let $\widehat H_\Lambda$ be the regularized quantum Hamiltonian of the theory in the Fock space,
$\Lambda$ being the parameter of (ultraviolet and infrared) regularization, $\Lambda\to\infty$,
and let $\widehat H_\Lambda^{ct}$ be the regularized counterterm Hamiltonian.
Then the Faddeyev transform is a unitary operator $U_\Lambda$
divergent as $\Lambda\to\infty$, such that the operator
\begin{equation}
U_\Lambda^{-1}(\widehat H_\Lambda+\widehat H_\Lambda^{ct})U_\Lambda=T_\Lambda
\end{equation}
has finite limit as $\Lambda\to\infty$. Such $U_\Lambda$ can be
constructed in each order of perturbation theory. Then the
(unbounded) operator $T=\lim_{\Lambda\to\infty}T_\Lambda$ can be
taken as the infinitesimal evolution operator in the Hilbert space
$\HH$. Similarly one constructs the action of other elements of
the Poincare group, using $U_\Lambda$.

Now, it is natural to assume that for each space-like surface $C$ and for each covariant Hamiltonian $H$ corresponding to
a first order deformation of $C$ [8,9], there exist operators $H_\Lambda^{ct}$ and $U_\Lambda(C)$
(divergent, in general, as $\Lambda\to\infty$) such that the operator $T_\Lambda=T_\Lambda(C)$ given by (1) with
$U_\Lambda(C)$ instead of $U_\Lambda$, has finite limit as $\Lambda\to\infty$. For example, this is true for free field.
Indeed, for free field $H_\Lambda^{ct}$ can be chosen to be a scalar, and $U_\Lambda$ can be chosen to be
an appropriate operator from the symplectic group acting on the Fock space. For a general space-like surface, the operator
$U_\Lambda(C)$ has no finite limit as $\Lambda\to\infty$, because the symplectic transformation corresponding to a
curved space-like surface is not in general implementable by a unitary operator in the Fock space, as shown in the paper
[5]. However, we need not $U_\Lambda$ to be convergent in our case.

It is natural to assume that the operators $T(C)$ form a flat connection $\nabla$ on the bundle of Hilbert spaces
$\HH_C$ corresponding to each $C$. This is motivated by the form of non-regularized covariant functional differential
Schrodinger equation for space-like surfaces [8,9]. We call $\nabla$ the {\it Tomonaga--Schwinger connection}.

Thus, we see how to overcome Torre--Varadarajan's paradox mentioned above: the Hilbert spaces $\HH_C$ for various $C$
cannot be canonically identified, except for the identification given by the Tomonaga--Schwinger connection. The only
additional identification between the spaces $\HH_C$ is given by the action of the Poincare group. Hence
we need not have unitary operators in one and the same Hilbert space corresponding to evolution between arbitrary curved
Cauchy surfaces.

Following the original approach of Tomonaga [1], one can identify the Hilbert space $\HH_C$ with the completion of
the space of expressions
\begin{equation}
\Psi|0\rangle=\sum_{n=0}^N\int_{C^n}\psi(s_1,\ldots,s_n)\varphi(x(s_1))\ldots\varphi(x(s_n)) ds_1\ldots ds_n|0\rangle,
\end{equation}
where $|0\rangle$ is the vacuum, $x=x(s)$ is a parameterization of the surface $C$, $\varphi(x(s))$ are the quantum field
operators at the points of the surface, $\psi(s_1,\ldots,s_n)$ is a smooth function on $C^n$ from the Schwartz space, and
the outer indices of the field are omitted.
Since $[\varphi(x),\varphi(y)]=0$ for $x-y$ a space-like vector, we have $[\varphi(x(s_i)),\varphi(x(s_j))]=0$, hence
$\Psi=\Psi(\varphi(s))$ can be considered as a polynomial functional
of a classical function (distribution) $\varphi(s)$ on $C$.

Let us
identify all the $\HH_C=\HH$ by means of the Tomonaga--Schwinger connection. Then the Poincare group action on $\HH_C$
yields a group of unitary operators in $\HH$, inverse to the Poincare group action in the WLSZ axioms.

Further, let us reconstruct the Green functions. For a smooth classical source function $j(x)$ with compact support
on $\R^{3+1}$, one can construct operators $\int j(x(s))\varphi(s)ds$ in $\HH_C$,
and for a one-parametric foliation $C(t)$ of space-time by space-like surfaces, consider the operator
\begin{equation}
T\exp\int j(x)\varphi(x)dx\eqdef T\exp\int_{-T}^T\int_{C(t)}j(x(s,t))\varphi(s)dsdt
\end{equation}
in $\HH$, where the number $T$ is so large that the support of the function $j(x)$ is situated between the surfaces
$C(-T)$ and $C(T)$. The generating functional of Green functions is given by
\begin{equation}
Z(j)=\langle 0|T\exp\int j(x)\varphi(x)dx|0\rangle.
\end{equation}
Note that $T\exp\int j(x)\varphi(x)dx$ is the evolution operator from $C(-T)$ to $C(T)$ of the connection
\begin{equation}
\nabla_j=\nabla+\int j(x(s))\varphi(s)ds
\end{equation}
on the bundle $\HH_C$, hence this evolution operator is correctly defined provided the connection $\nabla_j$ is flat.
The flatness of $\nabla_j$ means that the connection $\nabla$ is ``local'' (this is the definition of ``locality'').

Thus, we see that the mathematical definition of dynamical evolution on space-like surfaces in QFT
should be the following: it is a flat integrable connection $\nabla$ on the bundle of Hilbert spaces $\HH_C$
which are completions of the spaces of functionals (2) with respect to certain scalar products; this connection
should preserve the vacuum vectors, should be compatible with the group of symmetries of the theory
(e.~g. the Poincare group), and should be ``local'',
in the sense that for any smooth compact support function $j(x)$ on space-time, the connection $\nabla_j$
given by (5) is also flat (but can be integrable only as soon as $j(x)$ is considered as a formal perturbation
and the evolution operator is decomposed into the formal series with respect to $j$. This is necessary for construction of
Green functions).

\end{document}